\documentstyle[epsfig]{aipproc}
 
\def    \pl     #1#2#3{{\it Phys. Lett.} {\bf #1},~#3~(19#2)}
\def    \prl    #1#2#3{{\it Phys. Rev. Lett.} {\bf #1},~#3~(19#2)}

\def    \ibid   #1#2#3{{\it ibid.} {\bf #1},~#3~(19#2)}
\def    \hepph  #1 {{\tt hep-ph/#1}}
\def    \hepex  #1 {{\tt hep-ex/#1}}
\def\beq{\begin{equation}}
\def\eeq{\end{equation}}

\def\beqn{\begin{eqnarray}}
\def\eeqn{\end{eqnarray}}
\newcommand\sss{\scriptscriptstyle}

\begin{document}
\vspace*{-1in}
{\leftskip 9cm
\normalsize
\noindent   
\newline
hep-ph/9709427

}

\title{Heavy quark production: recent developments}

\author{Giovanni Ridolfi}
\address{INFN, Sezione di Genova\\Via Dodecaneso, 33, I-16146 Genova, Italy}

\maketitle
\begin{abstract}
I discuss some aspects of the comparison between QCD predictions
and experimental data in charm and bottom production.
\end{abstract}

Experimental information on the production of heavy-flavoured hadrons
is impressive. Charmed hadron production has been studied extensively
in fixed-target experiments, while colliders have provided
a detailed picture of the productions of $b$-flavoured hadrons.
Many aspects of the heavy-flavour physics have been studied in
$e^+e^-$ collisions at LEP.
Recently, the discovery of the top quark at the Tevatron has opened 
the way to a new set of tests of the heavy-quark production dynamics.
From the theoretical point of view, the production cross section
for heavy quarks has been computed in QCD up to next-to-leading
order, and detailed phenomenological analyses have been performed.
I will not attempt here a summary of the results obtained in
this field; a recent review of theoretical and experimental results and a
complete bibliography is given in \cite{Book}. In the following, I
will illustrate a few interesting issues which are currently under 
investigation.

I begin by considering first fixed-target hadro- and photo-production
of charmed mesons.
Comparisons between QCD predictions and data for the production of
charmed particles are clearly very difficult, because the value
of the charm quark mass, which sets the scale relevant
for the production mechanism, is very close to the region where
the applicability of perturbation theory becomes questionable. 
This said, one finds that
at fixed-target energies, $\sqrt{s}=20$ to 40~GeV,
the agreement between NLO QCD predictions and experimental data is 
in general 
satisfactory. Hadroproduction total cross sections are predicted
with an uncertainty of about one order of magnitude, due to the
large error coming from truncating the perturbative series at such low
energies. The prediction is also very sensitive to the value of the 
charm mass. The experimental uncertainties are much smaller, and
hadroproduction data are compatible with a charm mass of 1.5~GeV.
The situation is much more favourable in the case of photoproduction, where 
theoretical uncertainties are much smaller; in this case, unfortunaltely, some 
of the experimental data are incompatible with one another, and it will not be 
possible to use these data to put constraints on physical parameters
until these discrepancies are resolved.

\begin{figure}
\centerline{\epsfig{figure=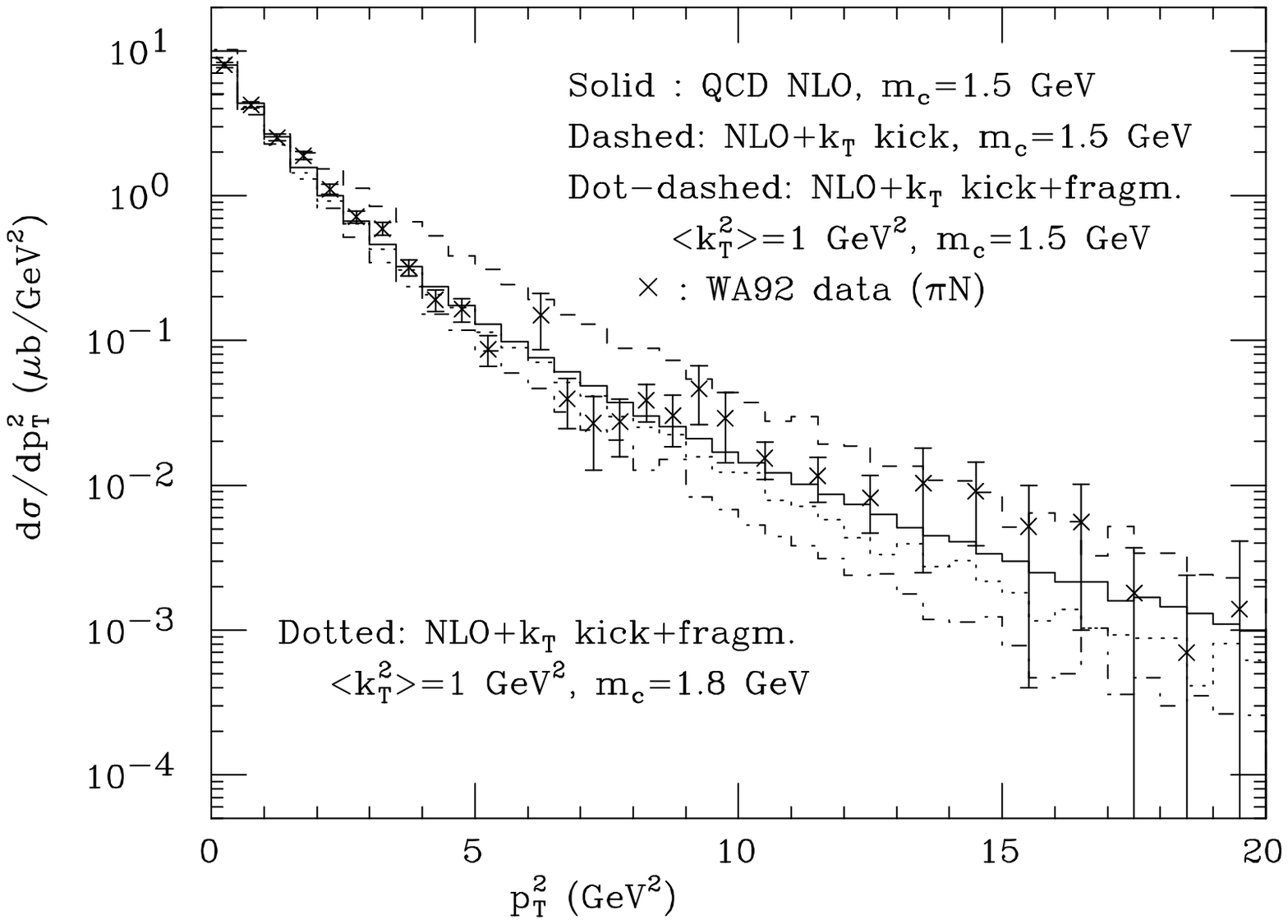,width=0.45\textwidth,clip=}
            \hspace{0.3cm}
            \epsfig{figure=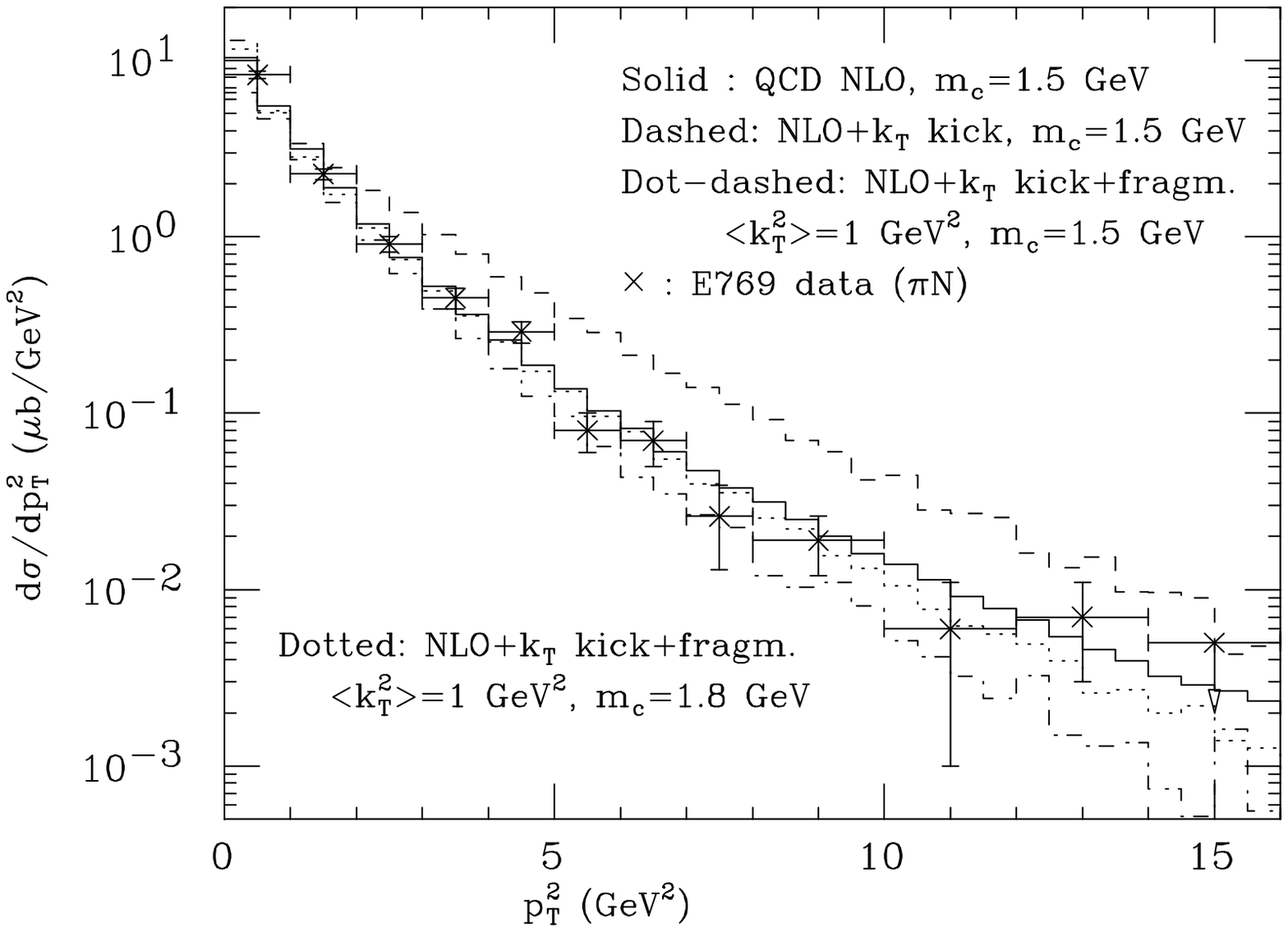,width=0.45\textwidth,clip=}}
\caption{}{ \label{hdrpt}\footnotesize
The single-inclusive $p_{\sss T}^2$ distribution measured by WA92 (left)
and E769 (right), compared to NLO QCD predictions,
with and without the inclusion of non-perturbative effects.}
\end{figure}

When distributions are considered, 
it is necessary to take into account also non-perturbative
effects, like the hadronization mechanism and the transverse momentum of 
partons in the colliding hadron beams.
A comparison between QCD predictions and data
for transverse quantities,
like the transverse momentum spectrum of charmed hadrons, shows a reasonable
qualitative agreement (see fig.~\ref{hdrpt}, where experimental
data from \cite{WA92pt,E769pt} and QCD predictions are shown).
Longitudinal quantities are
more difficult to predict reliably in a perturbative context,
and other non-perturbative effects (like for example color dragging)
must be taken into account. Also in this case, photoproduction data
show a better agreement with QCD.

\begin{figure}[htb]
\centerline{\epsfig{figure=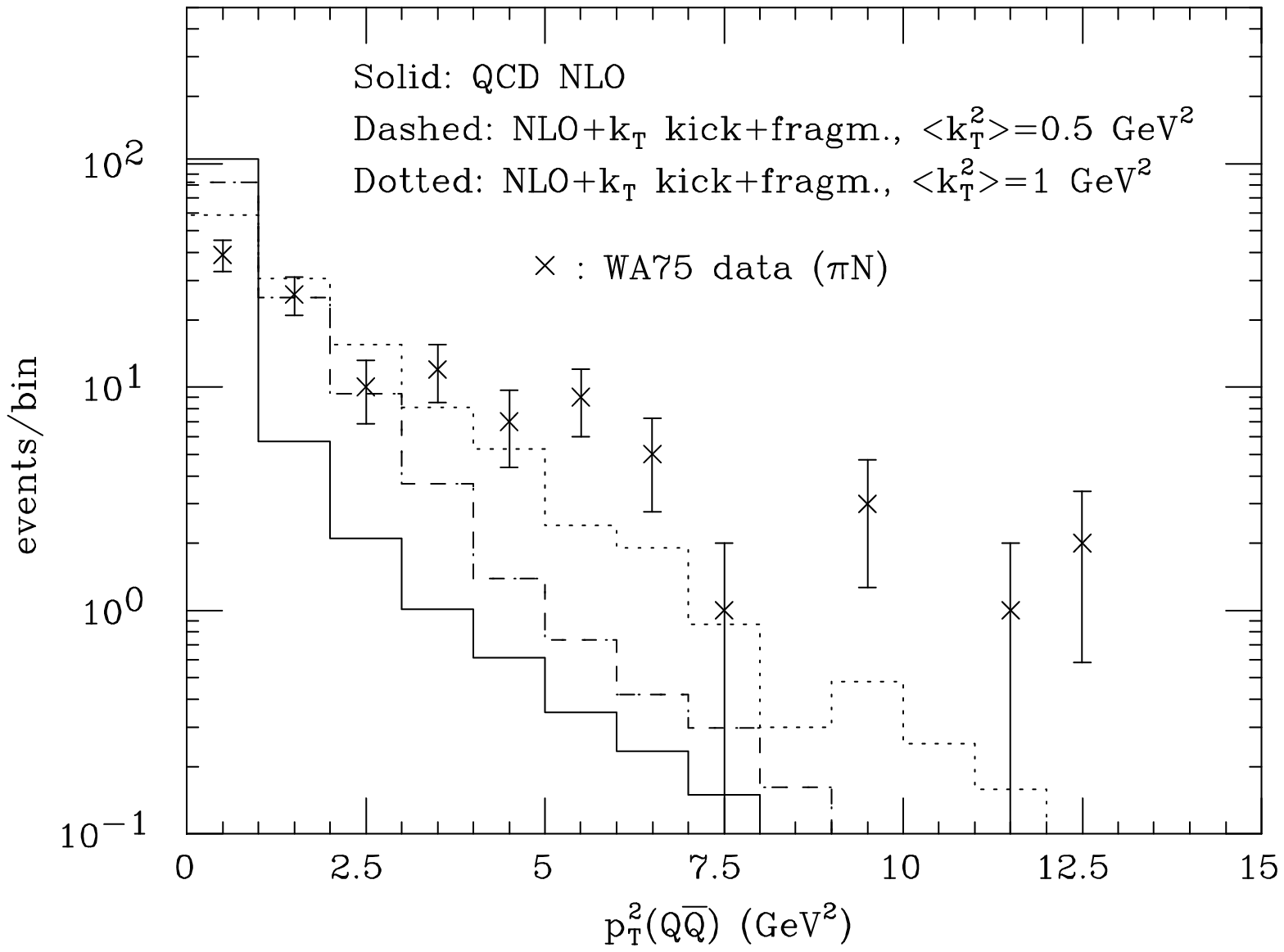,width=0.45\textwidth,clip=}
            \hspace{0.3cm}
            \epsfig{figure=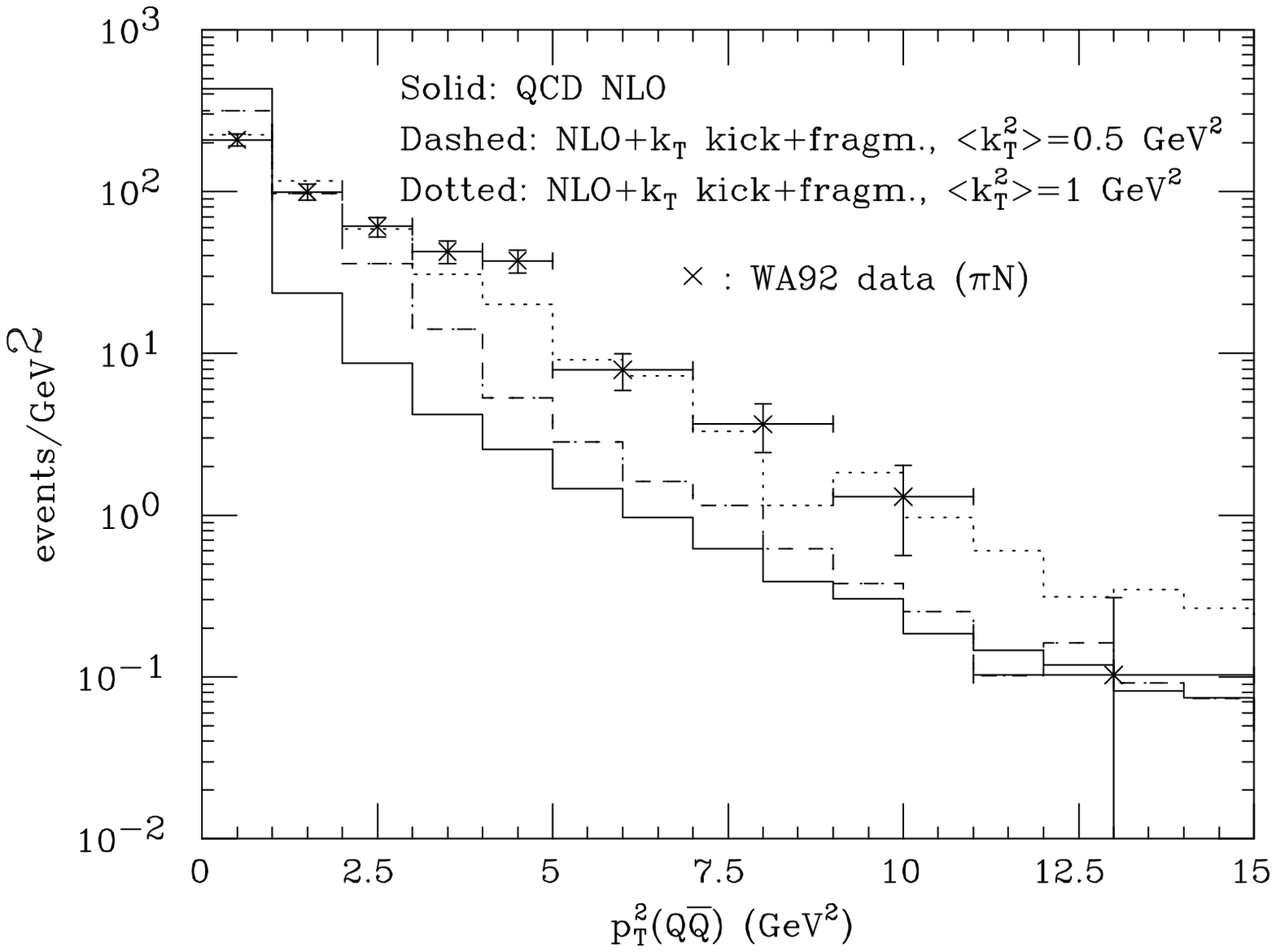,width=0.45\textwidth,clip=}}
\caption{}{\label{f:pt2qq}\footnotesize
NLO QCD result for the $p_{\sss T}^2(Q\overline{Q})$ supplemented
with an intrinsic transverse momentum for the incoming partons,
compared with the WA75 (left) and WA92 (right) data.
}
\end{figure}

Some measurements of double differential distributions, like the
azimuthal distance $\Delta\phi$ between the charmed hadrons, or the pair 
transverse momentum, have been performed by fixed-target 
experiments~\cite{WA75ptqq,WA92ptqq}.
The experimental results for azimuthal
$c\bar{c}$ correlations in hadron--hadron collisions show
a tendency to peak in the back-to-back
region $\Delta\phi=\pi$, but the peak is less pronounced than
the one predicted by perturbative QCD. The addition of an intrinsic transverse 
momentum of the incoming partons gives a satisfactory description of the
data.
The data on the $p_{\sss T}^2(Q\overline{Q})$ distribution
do not allow a unique interpretation. The theoretical prediction
is in rough agreement
with the WA92 measurement, and it is sizeably softer than the WA75 data,
as shown in fig.~\ref{f:pt2qq}

\begin{figure}
\centerline{\epsfig{figure=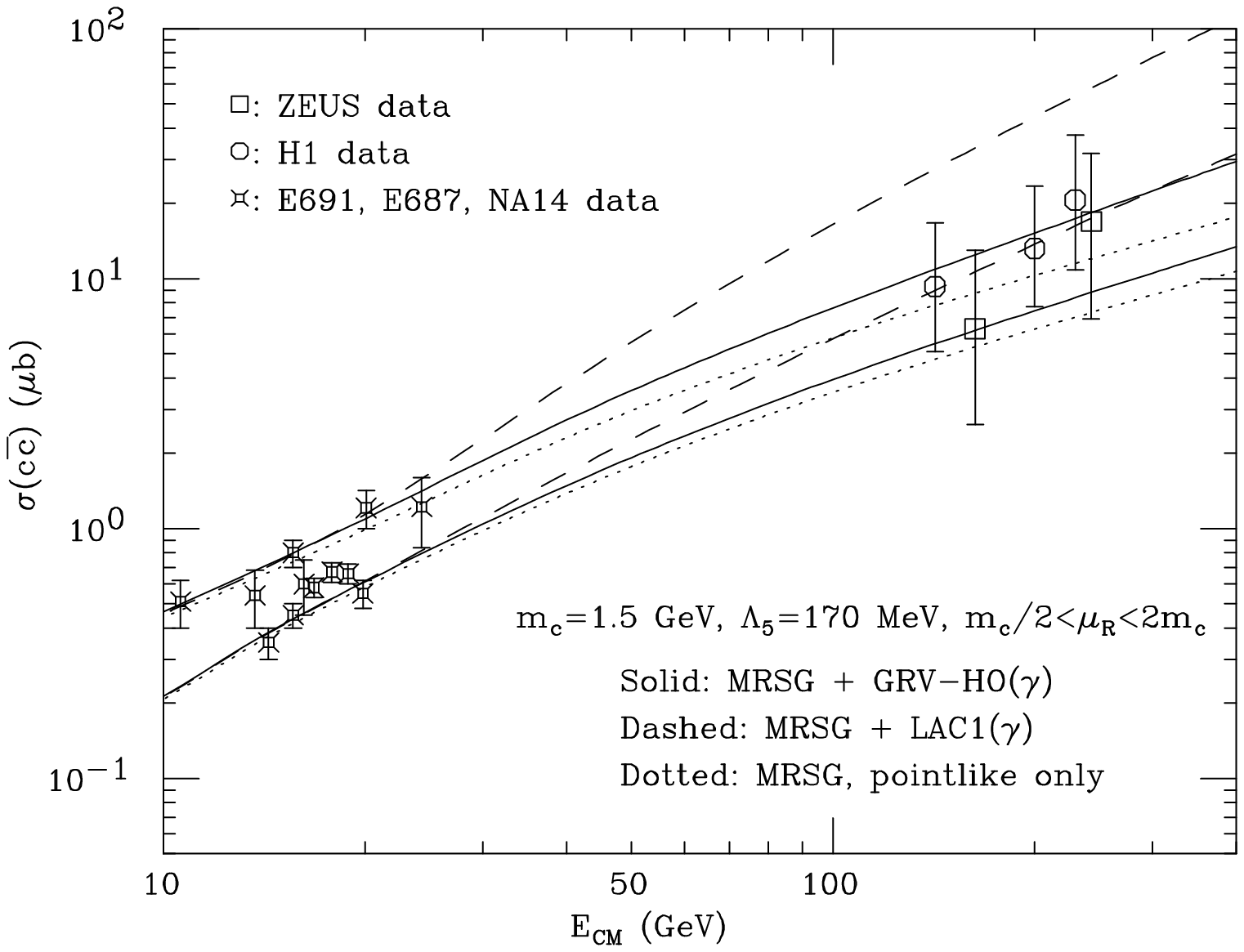,width=0.7\textwidth,clip=}}
\caption{}{ \label{f:sig_vs_ecm}\footnotesize
Total cross section for the photoproduction of $c\bar{c}$ pairs,
as a function of the $\gamma p$ center-of-mass energy:
next-to-leading order QCD predictions versus experimental results.}
\end{figure}

Experiments at the $ep$ collider HERA have also collected data on heavy quark 
production. At HERA, the center-of mass energy of the photon-proton
system is around 200~GeV; at these energies, the so-called hadronic
(or resolved) photon component of the cross section can become relevant.
In fig.~\ref{f:sig_vs_ecm} I present experimental data for the photoproduction
cross section of charm as a function of $\sqrt{s}$. Both fixed-target 
and HERA data are shown. Theoretical predictions, obtained with two different
parametrization of photon parton densities, are also displayed, together
with the corresponding uncertainties.
Observe that the HERA and fixed-target data are in good agreement
with theoretical expectations. Improvement of the statistics of the HERA data,
and the resolution of the discrepancies among fixed-target data
I mentioned above,
will probably allow to use total cross section data to put constraints on
photon parton densities.

HERA data on distributions and correlations are still at a
premature stage, and 
a significant comparison with theory will be possible when more statistics
will be collected. It is worthwhile mentioning that a direct measurement of
the gluon density in the proton using charm production is in principle
possible at HERA~\cite{fmnrglu}.

\begin{figure}[htb]
\centerline{\epsfig{figure=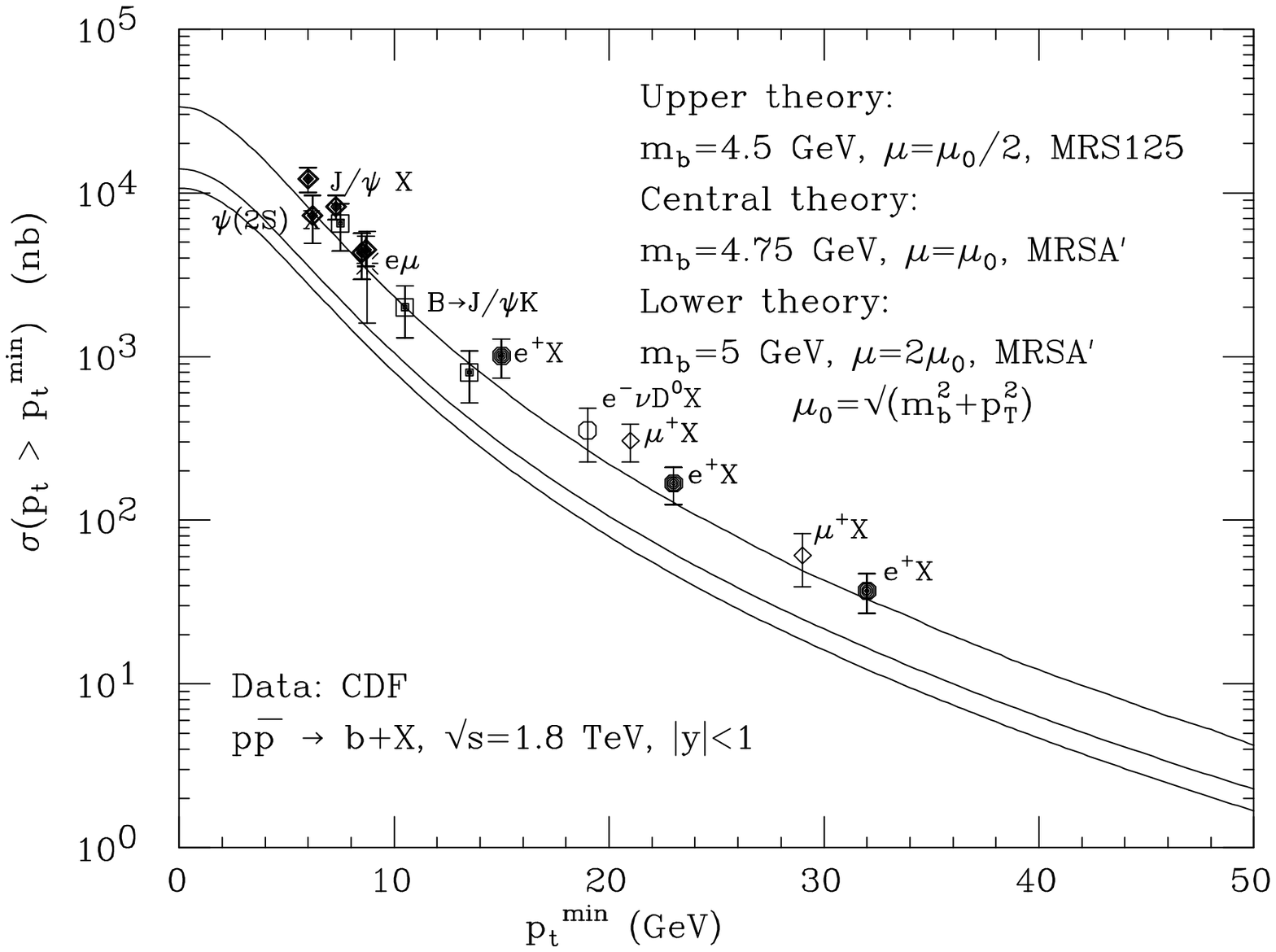,width=0.45\textwidth,clip=}
            \hspace{0.3cm}
            \epsfig{figure=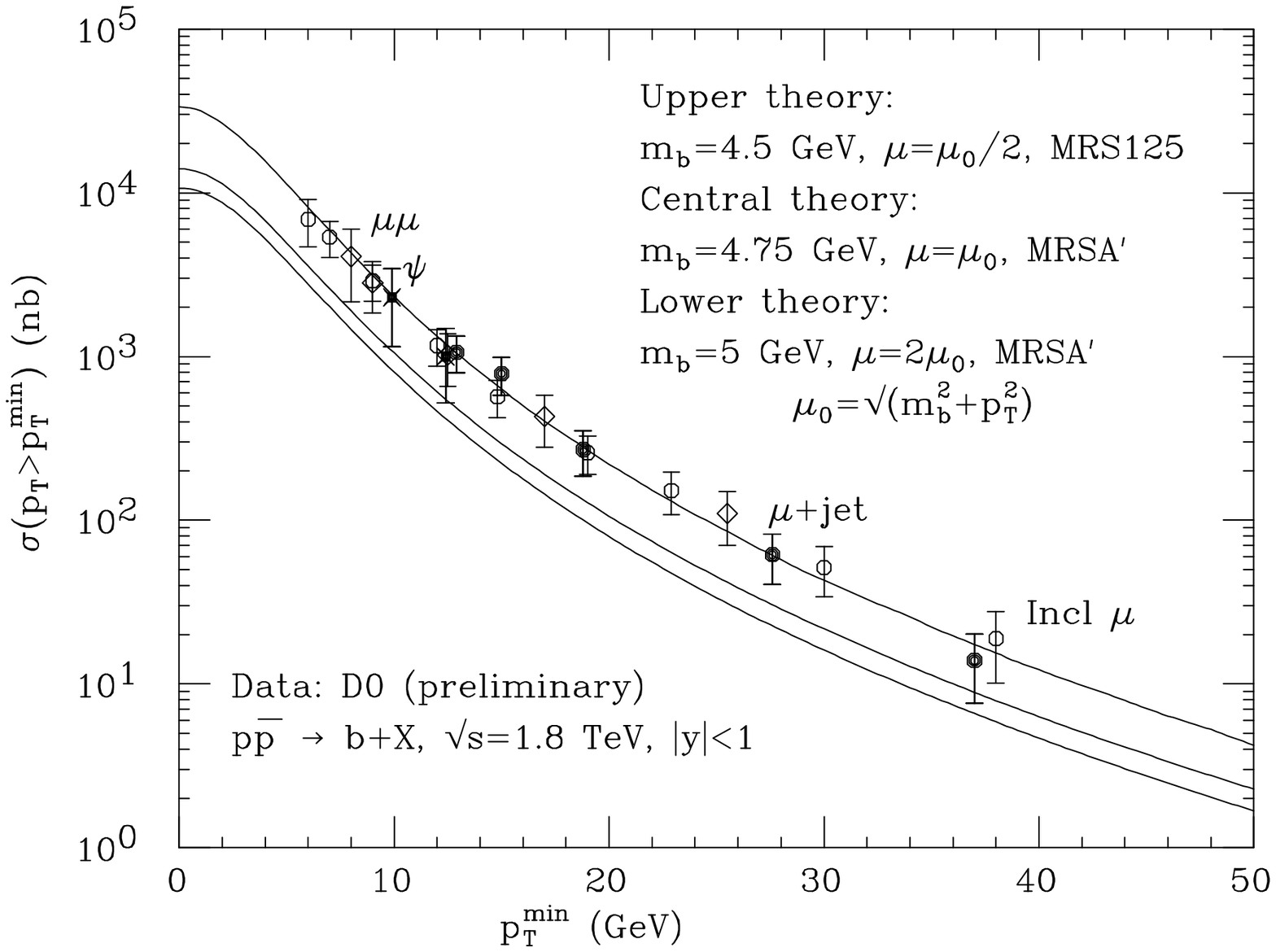,width=0.45\textwidth,clip=}}
\caption{}{\label{f:ptcoll}\footnotesize
CDF and D0 data on the integrated $b$-quark $p_{\sss T}$ distribution,
compared to the results
of NLO QCD.}
\end{figure}

Heavy-flavour production
in high-energy hadronic collisions offers a good opportunity for QCD tests.
For example, the $b$ quarks produced at large $p_{\sss T}$ can be studied in
perturbative
QCD with smaller contamination from non-perturbative effects, with respect
to charm.
The problem of the $b$ transverse momentum spectrum is a long-standing one.
The first measurements were performed by the UA1 collaboration, while
more recently the same distribution has been measured at the
Tevatron by CDF and D0, whose results are shown in fig.~\ref{f:ptcoll}.
The situation can be summarized as follows.
There is good agreement between the shape of the $b$-quark $p_{\sss T}$
distribution predicted by NLO QCD and that observed in the data for central
rapidities; although the data are higher by a factor of approximately 2
with respect to the theoretical prediction with the default choice
of parameters, extreme (although acceptable)
choices of $\Lambda^{\sss\overline{MS}}$ and of renormalization
and factorization scales
bring the theory in perfect agreement with the data of UA1 (not shown)
and D0, and within 30\% of the CDF measurements.
The choice of low values of the scales is favoured by studies of
higher-order logarithmic corrections.
The CDF measurements at 630 and 1800~GeV indicate that theory correctly
predicts the scaling of the differential $p_{\sss T}$ distribution
between 630 and 1800~GeV, a fact that had often been questioned in the past
and now finds strong support.

Forward production of $b$ quarks indicates a larger discrepancy between
theory and data, and more theoretical studies should be devoted to the
understanding of the
non-perturbative fragmentation function for heavy quarks.
The standard Peterson
parametrization may not be accurate enough for the description of
the hadroproduction data.

\bigskip
{\bf Acknowledgements:} All the results presented in this talk
have been obtained with Stefano Frixione,
Michelangelo Mangano and Paolo Nason; I would like to thank them
for this long and fruitful collaboration.

\end{document}